\documentclass[twocolumn,amssymb,footinbib, nobibnotes, aps, pra, reprint]{revtex4-1}

\usepackage{graphicx}
\graphicspath{{Figures/}}
\usepackage{bm}
\usepackage{braket} 
\usepackage{physics, mathtools}
\usepackage{hyperref}
\hypersetup{colorlinks,
	urlcolor=[rgb]{0,0.6,0.6},
	linkcolor=[rgb]{0,0.0,0.8}}

\renewcommand{\vec}{\mathbf}
\newcommand{\dG}{\delta\vec{\Gamma}}
\newcommand{\G}{\vec{\Gamma}}
\newcommand{\deltaG}{\delta\vec{\Gamma}}
\newcommand{\vecG}{\vec{\Gamma}}
\newcommand{\deltaQ}{\delta\vec{q}}
\newcommand{\deltaP}{\delta\vec{p}}
\newcommand{\Dt}{\Delta t}

\begin{document}	

\title{Estimating Lyapunov exponents in billiards}

\author{George Datseris}
\email{george.datseris@ds.mpg.de}
\affiliation{Max Planck Institute for Dynamics and Self-Organization}
\affiliation{Faculty of Physics, Georg-August-Universit\"at G\"ottingen}
\author{Lukas Hupe}
\thanks{GD and LH contributed equally to this work}
\affiliation{Max Planck Institute for Dynamics and Self-Organization}
\affiliation{Faculty of Physics, Georg-August-Universit\"at G\"ottingen}
\thanks{These two authors contributed equally}
\author{Ragnar Fleischmann}
\affiliation{Max Planck Institute for Dynamics and Self-Organization}
\affiliation{Faculty of Physics, Georg-August-Universit\"at G\"ottingen}

\date{\today}

\begin{abstract}
Dynamical billiards are paradigmatic examples of chaotic Hamiltonian dynamical systems with widespread applications in physics. We study how well their Lyapunov exponent, characterizing the chaotic dynamics, and its dependence on external parameters can be estimated from phase space volume arguments, with emphasis on billiards with mixed regular and chaotic phase spaces. We show that in the very diverse billiards considered here the leading contribution to the Lyapunov exponent is inversely proportional to the chaotic phase space volume, and subsequently discuss the generality of this relationship. We also extend the well established formalism by Dellago, Posch, and Hoover to calculate the Lyapunov exponents of billiards  to include external magnetic fields and provide a software implementation of it.

\end{abstract}

\maketitle

\textbf{From the foundations of statistical physics to transport properties of electronic devices, in many areas of physics \emph{billiard models} are an important tool for understanding complex dynamics. In a billiard model a point particle is moving freely (and frictionless) on a flat (or constantly curved) surface until it hits the boundary of the billiard where it is specularly reflected. Chaotic dynamics in the billiard are characterized by a positive \emph{Lyapunov exponent}, measuring how initially close trajectories separate exponentially fast. Obtaining its value so far usually requires detailed numerical simulations of the chaotic dynamics. In our paper we assess how well the Lyapunov exponent can be estimated from quite general considerations. Especially we study how parameter changes that vary the phase space structure of the billiard get reflected in the Lyapunov exponent. For example the application of an external magnetic field can force some trajectories in the billiard on closed cyclotron orbits. We show how the mere existence of such orbits varies the Lyapunov exponent of the chaotic dynamics through the phase space volume they occupy . The knowledge of this connection will be helpful to understand physical mechanisms in many systems like the magneto-transport in graphene nanostructures.
}

\section{Introduction}
Dynamical billiards are a well-studied class of dynamical systems, having applications in many different fields of physics. 
Besides playing a prominent role in ergodic theory~\cite{sinai_dynamical_1970, bunimovich_ergodic_1974, bunimovich_ergodic_1979}, billiards are important example systems for understanding quantum chaos~\cite{stockmann_quantum_1990, barnett_quantum_2007}, with practical applications e.g.~in modelling  optical microresonators for lasers~\cite{stone_nonlinear_2010-1, gmachl_high-power_1998} and room acoustics~\cite{koyanagi_application_2008}. 
Billiard-models have also been particularly successful in helping to understand transport properties of electronic nanostructures such as quantum dots and antidot super-lattices~\cite{datseris_observing_2017, Fleischmann1992, weiss_electron_1991, chernov_steady-state_1993, Yagi2015, Maier2017, Chen2016, bird2003, Beenakker2005, Ponomarenko2008}.

\begin{figure}[t]
	\centering
	\includegraphics[width = \columnwidth]{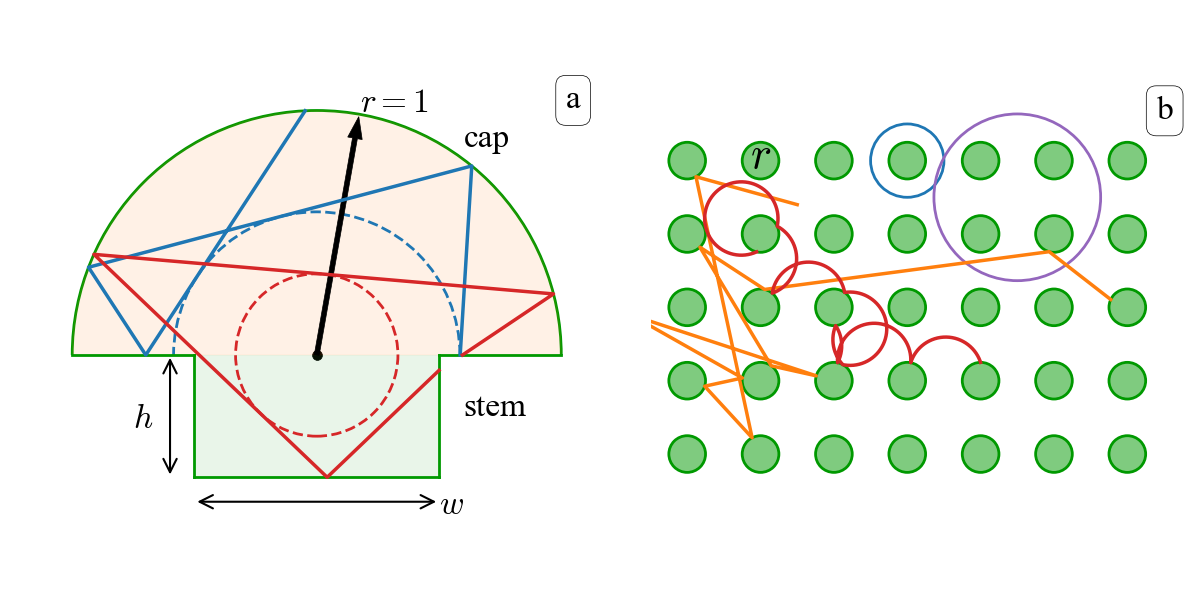}
	\caption{(a) A regular (blue) and chaotic (red) orbit in the mushroom billiard (MB), whose cap radius is a constant set to 1 and the stem has width $w$ and height $h$. The cap and the stem are separated with different background colors (orange, green). (b) chaotic orbits without (orange) and with (red) magnetic field and regular orbits (blue, purple) in the magnetic periodic Sinai billiard (MPSB) with disk radius $r$.}
	\label{fig:billiards}
\end{figure}

A billiard consists of a finite (or periodic) domain in which a point particle performs free flight with unit velocity. Upon collision with the boundary of the domain the particle (typically) is specularly reflected. In Fig.~\ref{fig:billiards} we are showing the two example billiards we will be considering in this paper: the  mushroom billiard (MB)~\cite{bunimovich_mushrooms_2001} and the periodic Sinai billiard~\cite{sinai_dynamical_1970} without (PSB) and with magnetic field (MPSB). 

An essential characterization of the chaotic dynamics of a billiard is of course provided by its Lyapunov exponents. (In this article we will study the Lyapunov exponents of the \emph{billiard flow} in the physical, continuous time, in contrast to those of the \emph{billiard map} in a discrete time that counts the number of collisions with the boundary). For a two dimensional billiard the Lyapunov exponents are four numbers $\lambda_{1 - 4}$ that measure how ``chaotic''  the billiard is, in terms of the average exponential rates of expansion (and contraction) of the phase space along certain characteristic directions. Due to the Hamiltonian nature of the dynamics, the Lyapunov exponents fulfil $\lambda_1 = -\lambda_4, \lambda_2 = \lambda_3 =0$. Therefore in the remainder of the text we will be only considering the largest exponent $\lambda \equiv \lambda_1$.  The fundamental mathematical properties of the Lyapunov exponents in billiards, including rigorous proofs of their existence, have been studied in literature and can e.g. be found in Ref.~\cite{chernov_entropy_1997, chernov_decay_2000, chernov_chaotic_book} and references therein.

Quantitative studies of the Lyapunov exponent in actual physical billiards are surprisingly rare however. A computational framework for calculating $\lambda$ in billiard systems was formulated by Dellago, Posch, and Hoover in Refs.~\cite{dellago_lyapunov_1995, dellago_lyapunov_1996} (to which we refer to as \textit{DPH framework} in the following, and which we will extend to the dynamics in magnetic fields). Alternative approaches are presented e.g in Refs.~\cite{chernov_new_1991, Garrido1997}. In the literature, Lyapunov exponents have been computed for the PSB on square~\cite{gaspard_chaotic_1995} and hexagonal~\cite{gaspard_chaotic_1995, dellago_lyapunov_1995} lattices, as well as for for the stadium billiard~\cite{dellago_lyapunov_1995, posch_simulation_2000}, which is related to the mushroom billiard. Furthermore, there are results for the magnetic elliptical billiard~\cite{meplan_tangent_1993} and the inverse magnetic stadium~\cite{voros_tunable_2003}.

All these quantitative calculations rely on detailed numerical simulations of the complex billiard dynamics. In this paper, however, we want to follow a different approach exploring approximate expressions for the parameter dependence of the Lyapunov exponents in some paradigmatic cases, especially of billiards with \emph{mixed} phase space, where regions of regular and chaotic dynamics coexist. Our work is motivated by a recent study that has shown that magnetoresistance measurements in graphene and semiconductor nanostructures directly reflect the parameter dependence of the chaotic phase space volume~\cite{datseris_observing_2017}. This is due to the fact that characteristic transport times in the chaotic sea are fundamentally linked to the respective volume of the chaotic phase space in the corresponding billiards. Namely, for the magnetic periodic Sinai billiard (MPSB) in Fig.~\ref{fig:billiards}b it was analytically shown that the mean collision time $\kappa(B)$ between successive collisions with the discs (of radius $r$) as a function of an applied external magnetic field $B$ is equal to the varying chaotic phase space portion $g_c(B)$ times the value of $\kappa$ at zero magnetic field~\cite{datseris_observing_2017}, i.e.
\begin{equation}
\kappa(B) = g_c(B) \times \kappa(0) = g_c(B) \times \frac{1 - \pi r^2}{2r}.
\label{eq:datseris_proof}
\end{equation}
(The value of $\kappa(0)$ is obtained from Eq.~\eqref{eq:universal} below.) For the convenience of the reader we replicate the proof of eq.~\eqref{eq:datseris_proof} from~\cite{datseris_observing_2017}, which uses Kac's lemma~\cite{kac_notion_1947, mackay_transport_1994,meiss_average_1997}, in Appendix~\ref{ap:sinaiproof}.

The Lyapunov exponent in billiards is also linked to mean collision times as the following back-of-the-envelope calculation motivates. Let us study the perturbation growth, i.e.~the exponential growth of the phase space distance $|\dG(t)|$ of two initially infinitesimally close by trajectories. The origin of the exponential perturbation growth and thus of chaos in billiards are collisions with curved boundaries~\cite{Bunimovich2018}. Assuming an average perturbation growth increase of $C$ between collisions with curved boundaries and an average time $\kappa$ between such collisions, one would expect a perturbation growth of $|\dG (t)| \approx C^{t/\kappa}|\dG (0)|$. This means for the Lyapunov exponent of the ergodic component of  phase space (see definition in Eq.~\eqref{eq:definition_lyapunov} below) we expect 
\begin{equation}
    \lambda \approx \frac{\log(C)}{\kappa}.
    \label{eq:back-of-envelope}
\end{equation}

In general, billiards have non-curved boundaries as well as curved ones. The mean collision time $\tau$ between two consecutive collisions with any parts of the billiard boundary  is known analytically for \emph{any} billiard and is given by
\begin{equation}
    \tau = \frac{\pi |Q|}{|\partial Q|}
    \label{eq:universal}
\end{equation}
where $|Q|$ is the area of the billiard and $|\partial Q|$ the total length of its boundary~\cite{chernov_entropy_1997}. Since eq.~\eqref{eq:universal} is averaged over the entire boundary of the billiard it includes contributions from both chaotic and regular components (if any). Also notice that a formula similar to eq.~\eqref{eq:universal} exists for billiards of any dimension, see~\cite{chernov_entropy_1997}. 

The starting point our work is the observation that in billiards the mean collision time between curved boundaries  $\kappa$ (a more precise definition will be given in section~\ref{sec:pertgrowth}) is fundamentally linked to the chaotic phase space volume $V_\mathrm{CH}$ by Kac's lemma~\cite{kac_notion_1947, mackay_transport_1994,meiss_average_1997}. Therefore also the Lyapunov exponent is linked to the chaotic phase space volume, and the aim of this paper is to explore how far considerations like Eq.~\eqref{eq:datseris_proof} allow us to estimate the parameter dependence of the Lyapunov exponent in billiard systems. To this end we will analyze the contributions to the approximate expression~\eqref{eq:back-of-envelope} and compare it with detailed numerical simulations. In sec.~\ref{sec:dellago} we provide the basic framework we will use for computing $\lambda$, as well as apply the aforementioned back-of-the-envelope calculation to realistic perturbation growth. Following in sec.~\ref{sec:results} we present our results about the periodic Sinai billiard and the mushroom billiard. We conclude by discussing the generality of our approach, while presenting one additional billiard with mixed phase space, the inverse stadium billiard, which has been studied by V\"or\"os et al.~\cite{voros_tunable_2003}.

\section{Lyapunov exponents in billiards}
\label{sec:dellago}
In this section we first give a brief overview of the DPH framework~\cite{dellago_lyapunov_1995, dellago_lyapunov_1996} for numerically computing $\lambda$, reciting the equations that will be relevant for our study. We will then extend the framework to motion in an external magnetic field. In the following we will assume that the point particle in the billiard has unit mass and momentum and velocity are the same.

The (maximum) Lyapunov exponent is defined based on the evolution of the four-dimensional perturbation vector $\delta \vec{\Gamma} = (\delta \vec{q}, \delta\vec{p})^T$ as
\begin{equation}
    \lambda_{\G(0), \dG(0)} = \lim_{t\to \infty}\frac{1}{t}\log \frac{|\dG(t)|}{|\dG(0)|}
    \label{eq:definition_lyapunov}
\end{equation}
with $\dG$ evolving according to the evolution equations in tangent space, $\delta\dot{\vec{\Gamma}} = J(\vec{\Gamma(t))}\cdot \delta\vec{\Gamma}$ where $J$ is the Jacobian matrix of the equations of motion. For all $\G (0)$ inside an ergodic component of phase space the value of $\lambda$ does not depend on the initial condition.

\subsection{Without magnetic field}

The time evolution in tangent space for a particle moving in a straight line is
\begin{equation}
  \label{eq:evolution_free}
  \dG(t) = \left(\begin{array}{c|c}
                        \mathbb{I}_{2\times2}&t\cdot\mathbb{I}_{2\times2}\\\hline
                        0_{2\times2}&\mathbb{I}_{2\times2}
                    \end{array}\right)\;\dG (0).
\end{equation}
At discrete time points Eq.~\eqref{eq:evolution_free} is interrupted by collisions with the boundary and the perturbation vector changes discontinuously. The perturbation vector just after the collision (we'll use $'$ to label quantities right after the collision) is derived from the one just before the collision as~\cite{dellago_lyapunov_1996}
\begin{equation}
    \label{eq:evolution_disk}
    \deltaG' = \left(\begin{array}{l}
        \deltaQ - 2\left(\deltaQ \cdot \vec{n}\right)\vec{n}\\
        \deltaP - 2\left(\deltaP \cdot \vec{n}\right)\vec{n} - 2\gamma_r \;
            \tfrac{\left(\deltaQ\cdot \vec{e}\right)}{\cos\phi}\;\vec{e}'\\
    \end{array}\right)
\end{equation}
for a collision with a boundary segment of curvature $\gamma_r$. The two types of boundaries we will encounter in this work are straight walls and the boundaries of circular discs. For a straight wall section $\gamma_r=0$ and for a disc of radius $r$ we have $\gamma_r=\pm\frac{1}{r}$, with $-$ for collisions happening from the inside of the disc (as in the MB) and $+$ otherwise (as in the PSB). Here $\vec{n}$ denotes the normal vector of the boundary segment at the collision point $\vec{q}$, and $\phi$ is the angle of incidence (measured with respect to the normal to the surface).  The vectors $\vec{e}, \vec{e}'$ are unit vectors orthogonal to the incident and reflected momenta $\vec{p}$ and $\vec{p}'$ respectively (for more details see~\cite{dellago_lyapunov_1996}). 

Notice that a collision with a straight wall does not change the norm of $\dG$ because both the coordinate and the velocity components are reflected specularly.

\begin{figure*}[t!]
    \centering
    \includegraphics[width=\textwidth]{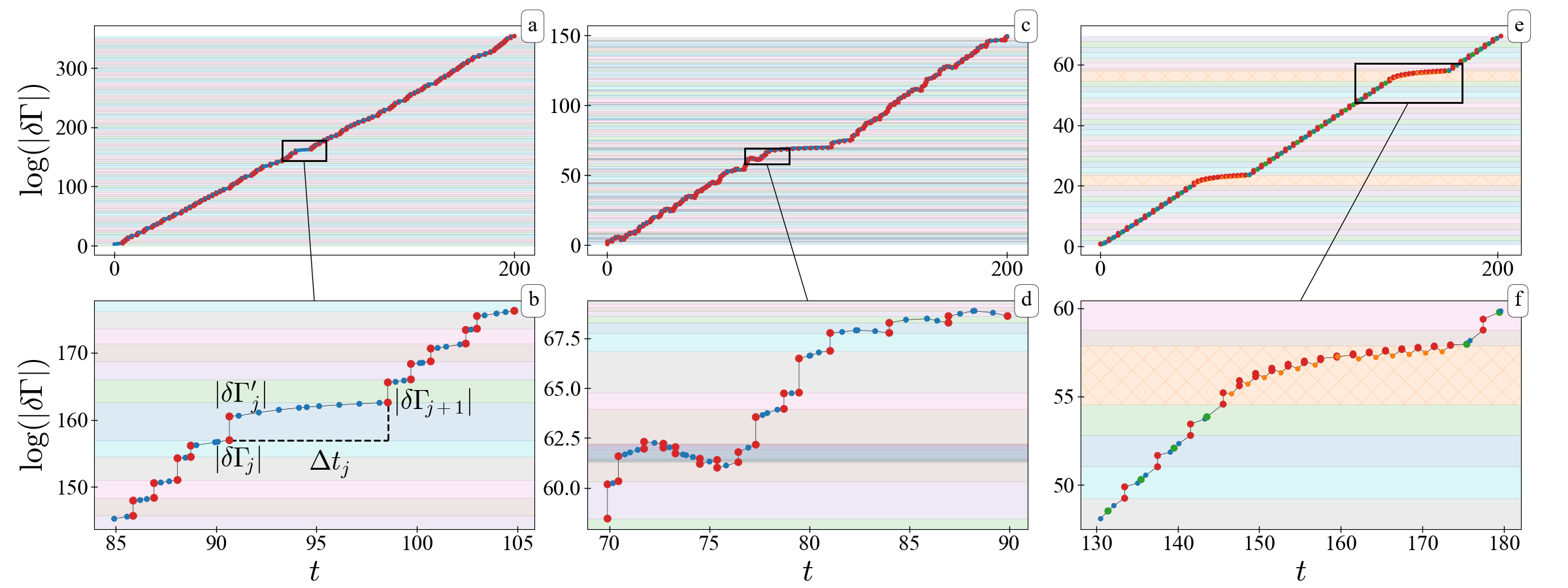}
    \caption{Typical time evolution of the logarithm of the norm of the perturbation vector $|\dG(t)|$ for the periodic Sinai billiard without (a) and with (c)  magnetic field ($B=1$) and for the mushroom billiard (e) (without magnetic field). Zoom-ins are below each panel. For (a-d) red markers mean collision with the disc while blue markers mean ``collision'' with the periodic walls (not a true collision, but a way of recording the value of $|\dG|$). For (e-f) red is collision with cap head, orange with cap walls, blue with stem sides and green with stem bottom. The coloured background stripes denote the \emph{elementary growth segments} discussed in sec.~\ref{sec:pertgrowth} (random colors are used) with hatched orange color used for the laminar episodes (see sec.~\ref{sec:mushroom}).}
    \label{fig:pertgrowth}
\end{figure*}

\subsection{With magnetic field}
We now extend the DPH formalism for a particle experiencing a magnetic field perpendicular to the billiard plane. In this section we present only the final expressions. The full calculations are presented in appendix~\ref{ap:magnetic}.

The magnetic field is uniform, with value $B$ (positive means counter-clockwise rotation). The free evolution of the perturbation vector $\dG(t)$ in the presence of a perpendicular magnetic field is given by
\begin{align}
  & \dG(t) = \mathbb{B} \cdot \dG(t_0), 
  \label{eq:evolution_magnetic} \\
  & \mathbb{B} = \left(
  \begin{array}{c|cc}
    \mathbb{I}_{2\times2} & 
             \begin{array}{c}
               \rho \sin(\omega t)\\
               -\rho \left(\cos(\omega t) - 1\right)
             \end{array} &
                           \begin{array}{c}
                             \rho \left(\cos(\omega t) - 1\right)\\
                             \rho \sin(\omega t) 
                           \end{array} \\ \hline
    0_{2\times2}&
             \begin{array}{c}
               \cos(\omega t)\\
               \sin(\omega t)
             \end{array} &
                           \begin{array}{c}
                             -\sin(\omega t)\\
                             \cos(\omega t)
                           \end{array}
  \end{array}
  \right) \nonumber
\end{align}
with the cyclotron frequency $\omega = 2B$ and the cyclotron radius $\rho = 1/\omega$. As already mentioned in the introduction, in the billiard the particle always moves with unit velocity by convention. The expressions that give the discontinuous change of the perturbation vector at a collision with a wall or disc are
\begin{align}
    \dG'  =& \pmqty{
    \delta\vec{q} - 2\left(\delta\vec{q}\cdot\vec{n}\right)\vec{n}\\
    \delta\vec{p} - 2\left(\delta\vec{p}\cdot\vec{n}\right)\vec{n} - 2\gamma_r\;
    \tfrac{\left<\delta\vec{q},\vec{e}\right>}{\cos \phi}\;\vec{e}'\\
  } - \nonumber \\
  & \omega\,\frac{\left(\delta\vec{q}\cdot\vec{n}\right)}{\left(\vec{p}\cdot\vec{n}\right)}
  \pmqty{0\\ \mathbb{S}\cdot \vec{p}}
  \label{eq:evolution_magnetic_disk} \\
  \mathbb{S} = & 2
  \begin{pmatrix}
    - 2n_1n_2 & n_1^2 - n_2^2\\
    n_1^2-n_2^2 & 2n_1n_2
  \end{pmatrix},\quad \vec{n} = \begin{pmatrix} n_1 \\ n_2 \end{pmatrix} \nonumber
\end{align}
where $\gamma_r$ again is the curvature of the wall segment ($0$ for a straight wall, $\pm\tfrac{1}{r}$ for a disc).

\subsection{The ``toy model''}
\label{sec:pertgrowth}
Before finding an approximate expression for the value of $\lambda$ in our model systems, it is worthwhile to get an impression of how the norm of the perturbation vector evolves with time. In Fig.~\ref{fig:pertgrowth} we show typical plots for the three different billiards. We computed the perturbation growth using the DPH framework, sampling the perturbation vector immediately before and after every collision to resolve the instantaneous jumps. As the DPH evolution is linear, in the actual numerical simulations we renormalized the perturbation vector after sampling to prevent numerical error due to the rapid perturbation growth.

Let us first examine Fig.~\ref{fig:pertgrowth}(a, b). We see that the norm of the perturbation vector changes in two ways. Let the $j$-th collision with a disc happen at time $t_j = \sum_{i=0}^{j-1} \Delta t_i=t_{j-1}+\Delta t_{j-1}$. There a discontinuous change of the perturbation vector norm happens, so that $|\dG'_j| = a_j|\dG_j|$ (in general $a_j$ is a function of $\dG_j$). The collision event is followed by a time-interval $\Delta t_j$, in which the perturbation norm changes continuously because there are no collisions with curved boundaries. Just before the next collision with a disc the perturbation norm takes the value $|\dG_{j+1}|$. In the following we will refer to these repeated segments of the growth curve as \emph{elementary growth segments}, starting with one collision event with a disc and ending just before the next one. In general it is the segment of the perturbation growth curve between successive dispersing or defocussing collisions which are the origin of chaos in billiards~\cite{Bunimovich2018}. An elementary growth segment thus reflects the perturbation growth during an ``effective free path'' as defined by Bunimovich~\cite{Bunimovich2018}.

A crucial simplification we do in deriving an approximate expression for the Lyapunov exponent will be to express the perturbation growth in the time-interval  $\Delta t_j$ as a function $z(\Delta t_j)$ of the interval length.
The actual precise value $|\dG_{j+1}/\dG'_{j}|$ is not a simple scalar function of the time interval since for example in the case of the PSB we can obtain from Eq.~\eqref{eq:evolution_free} 
\begin{equation}
    |\dG_{j+1}(\Dt)| = \sqrt{(\deltaQ_j' + \deltaP_j'  \Dt)^2 + (\deltaP_j')^2}
    \label{eq:noB_increase}
\end{equation}
(due to the linearity of the equations of motion of the tangent space we can assume a norm of $|\dG_j'| = 1$ in Eq.~\eqref{eq:noB_increase}). Eq.~\eqref{eq:noB_increase} depends on the initial orientations of both the momentum and position deviation vectors and thus is not a function of just $\Delta t$.

We will show, however, by analysing numerical data, that a reasonable approximation can be obtained  by assuming that such a function $z(\Delta t)$ exists. Notice that this assumption regards only the existence of $z$. The functional form and its complexity can be completely arbitrary (and in fact in the following we have three different versions of $z$ for the different billiards). 
For each elementary growth segment we thus write
\begin{equation}
    |\dG_{j+1}| =  z(\Delta t_j) \times a_j \times |\dG_j| = z(\Delta t_j) \times |\dG_j'| .
    \label{eq:toy_initial}
\end{equation}
We then recursively apply Eq.~\eqref{eq:toy_initial} to get
\begin{align}
    |\dG_{n}| &= \prod_{i=0}^{n-1} a_i z(\Delta t_i) |\dG(0)| \quad \Rightarrow \nonumber \\
    \log\left(|\dG_{n}|\right) &= \sum_{i=0}^{n-1} \log(a_i) + \log(z(\Delta t_i))
\end{align}
(using $|\dG(0)| = 1$) and with $T_n = \sum^{n-1}_{i=0} \Delta t_i$ we use Eq.~\eqref{eq:definition_lyapunov} to write  $\lambda = \lim_{n\to\infty} \log\left(|\dG_{n}|\right)/T_n$. 
The quotient of the infinite sums is the same as the ratio of the average over all unit cells (denoted by $\Braket{\cdot}$), i.e.
\begin{align}
    \lambda &\approx \frac{1}{\kappa}\left(\Braket{\log(a)} + \Braket{\log(z(\Delta t))} \right) 
    \label{eq:toymodel} \\
    \kappa & \equiv \Braket{\Delta t}.
\end{align}
Averaging over the unit cells implicitly assumes the ergodicity.
Notice also that in some billiards there could be several ergodic chaotic components that are separated from each other. In such cases the above process has to be applied to each component separately, since each component has its own exponent $\lambda$.
For the billiards considered here we have found that furthermore $\Braket{\log(z(\Delta t))} = \log(z(\Braket{\Delta t}))= \log(z( \kappa))$ is a good approximation that we will use.
This approximation is valid when the standard deviation of $\Delta t$ is small (compared to its mean).

In the remainder of the text we will call Eq.~\eqref{eq:toymodel} the ``toy model''. It is the more detailed version of the back-of-the-envelope calculation given in the introduction. In the following sections we will apply this toy model to specific billiards and see how well it approximates the Lyapunov exponent and its parameter dependence.

\subsection{Software}
All numerical computations presented in this paper were performed with an open source software to simulate billiards, \textbf{DynamicalBilliards.jl}~\cite{datseris_dynamicalbilliards.jl_2017}. In this paper we extend the DPH framework to magnetic fields. We also implemented this extension in the software (which previously only included the non-magnetic case).
All code we used for this paper, including all code to replicate the figures we show here, is publicly available on GitHub: \url{https://github.com/Datseris/arXiv_1904.05108}.

\section{Results}
\label{sec:results}
\subsection{Periodic Sinai billiard}
We start our analysis with the PSB because it is the simplest case and we are able to give a fully analytical expression for the Lyapunov exponent in the simplified toy model. We note that in the absence of a magnetic field the PSB is ergodic and its phase space is not mixed. Nevertheless, it will serve as a pedagogic example of how the toy model approximates the Lyapunov exponent.

\begin{figure}[t!]
    \centering
    \includegraphics[width=\columnwidth]{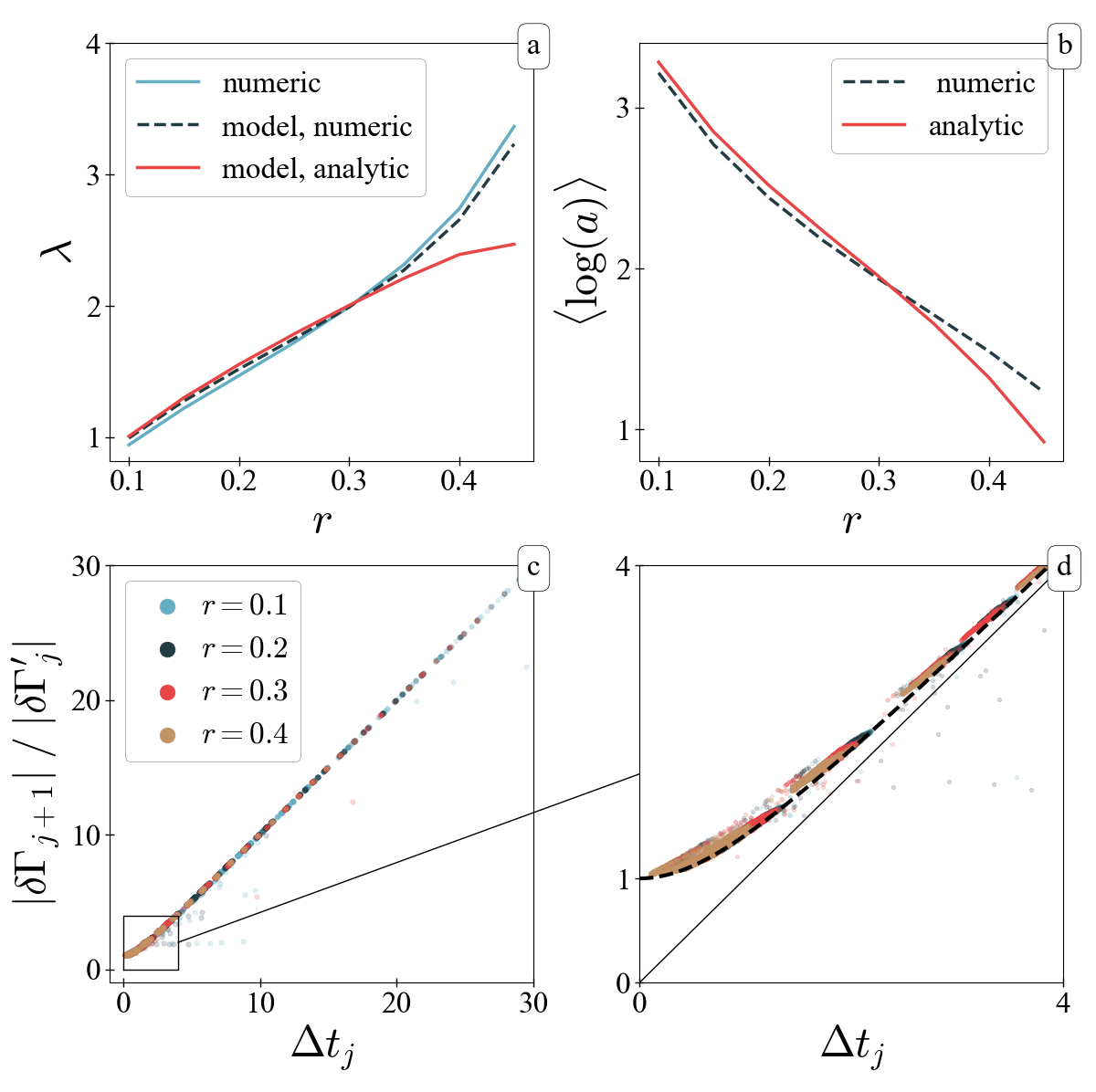}
    \caption{(a) Lyapunov exponent of the periodic Sinai billiard for different radii, compared with the toy model. The dashed curve obtains $\Braket{\log(a)}$ by numeric average while the red curve uses Eq.~\eqref{eq:psb_loga}. The blue curve is using the DPH framework. (b) Average value of $\Braket{\log(a)}$ used in panel (a). (c) Perturbation norm increase during the free flight part. In the zoom in (d) we also plot the curve $\sqrt{1+\Delta t^2}$ as a dashed line.}
    \label{fig:psb}
\end{figure}

For the PSB $\tau = \kappa$ and the value of $\tau$ is known from Eq.~\eqref{eq:universal}
\begin{equation}
    \kappa_{\text{PSB}} = \frac{1 - \pi r^2}{2r}.
    \label{eq:tau_psb}
\end{equation}
An approximation for $z(\Delta t)$ is easily found as well from Eq.~\eqref{eq:noB_increase}, namely
\begin{equation}
    z_{\text{PSB}}(t) \approx \sqrt{1 + t^2}
    \label{eq:g_psb}
\end{equation}
which uses the assumption that after a collision with a disc the momentum contribution to the perturbation vector is much larger than the position contribution, i.e. $|\deltaP'|\gg |\deltaQ'|$. Numerical calculations show that this approximation is valid for small enough radii (see Fig.~\ref{fig:psb}). The instantaneous change factor $\Braket{\log(a)}$ is rather large for all but very large disc radii. Also as seen in panel (c) and its inset (d), Eq.~\eqref{eq:g_psb} almost perfectly approximates the perturbation norm increase during the free flight part.

We still need an approximate expression for ${a}$, the instantaneous change factor, which we can derive from Eq.~\eqref{eq:evolution_disk}. Since the norm of the position deviation does not change at the collision, we focus on the momentum deviation $\delta \vec{p}' = \delta \vec{p}^{(r)} - \frac{2}{r}\frac{\delta \vec{q} \cdot \vec{e}}{\cos(\phi)} \vec{e}'$ with $\delta \vec{p}^{(r)} = \deltaP - 2\left(\deltaP \cdot \vec{n}\right)\vec{n}$ (if not explicitly written otherwise all quantities in this paragraph have a collision-time index $j$, which we suppress to make the symbols simpler). We define $A=\frac{2}{r}\frac{\delta \vec{q} \cdot \vec{e}}{\cos(\phi)}$ and carrying out the calculations leads to
\begin{equation}
|\dG'|/|\dG| = a =
\sqrt{1+ A^2 - 2 A (\vec{e}' \cdot \delta\vec{p})}
\label{eq:pre_a}
\end{equation}
again using the assumption that $|\dG_j| = 1$ (and recall that $|\delta \vec{p}^{(r)}| = |\deltaP|, |\deltaQ'|=|\deltaQ|$). 

We now need to average $\Braket{\log(a)}$. We start our approximation by replacing the inner product $(\deltaQ\cdot\vec{e})$ by an averaged quantity $b(r)$ (we show below how $b$ depends on $r$). It is expected that perturbations will grow and orient themselves perpendicular to the particle's direction of motion. Since $\vec{e}$ is a unit vector perpendicular to the particle's momentum and thus parallel to $\deltaQ$, this means that $ \Braket{|(\deltaQ\cdot\vec{e})|/|\dG|} = \Braket{|\deltaQ|/|\dG|}=b(r)$. 

Which portion of $|\dG|$ is contained in $|\deltaQ|$ is answered based on how perturbations evolve during the free flight part. Starting from the $j$-th collision the perturbations evolve for time $\Delta t_j$.
Using Eq.~\eqref{eq:noB_increase} (and assuming that the cross-terms $\deltaP \cdot \deltaQ$ will drop out in the averaging) we obtain
\begin{equation}
    \frac{|\deltaQ_{j+1}|}{|\dG_{j+1}|} = \frac{\sqrt{\frac{|\deltaQ_j'|^2}{|\deltaP_j'|^2} + \Delta t_j^2}}{\sqrt{\frac{|\deltaQ_j'|^2}{|\deltaP_j'|^2}+ 1 + \Delta t_j^2}}.
    \label{eq:b_pre}
\end{equation}

To analytically resolve eq.~\eqref{eq:b_pre} we use the same assumption as in eq.~\eqref{eq:g_psb},  $|\deltaP'|\gg |\deltaQ'|$. We then average, replacing $\Delta t$ by $\kappa$, which leads to
\begin{equation}
    b(r) = \sqrt{\frac{\kappa_{\text{PSB}}^2}{1 + \kappa_{\text{PSB}}^2}}.
    \label{eq:b_factor}
\end{equation}

We discard the term $2 A (\vec{e'}\cdot \delta\vec{p})$ in Eq.~\eqref{eq:pre_a} again assuming that the inner product averages to 0. Now the only variable left to average over is $\phi$, the angle with respect to the normal vector. This is distributed in $[-\pi/2, \pi/2]$ with probability distribution of $\cos(\phi)/2$. Therefore
\begin{align}
    \Braket{\log(a)} & =  \int_{-\frac{\pi}{2}}^{\frac{\pi}{2}} \log\left(\sqrt{1+\left(\frac{2b}{r\cos(\phi)}\right)^2}\right) \frac{\cos(\phi)}{2}\,d\phi \nonumber \\
     & = \frac{\text{csch}^{-1}\left(\frac{2b}{r}\right) \sqrt{4b^2 +r^2} }{r}+\log \left(\frac{b}{r}\right)
    \label{eq:psb_loga}
\end{align}
with $\text{csch}^{-1}$ the inverse hyperbolic cosecant.

We then put eqs.~\eqref{eq:tau_psb}, \eqref{eq:g_psb}, \eqref{eq:b_factor} and \eqref{eq:psb_loga} into the toy model of Eq.~\eqref{eq:toymodel} and obtain an analytic approximation for the Lyapunov exponent
\begin{align}
    \lambda_\text{PSB}(r) = & \frac{2r}{1-\pi r^2}\left( 
    \frac{\text{csch}^{-1}\!\left(\frac{2b(r)}{r}\right) \sqrt{4b(r)^2 +r^2}  }{r}+ \nonumber \right. \\
    & \left. \log \left(\frac{b(r)}{r}\right) +\log\left( \sqrt{1 +  \left(\frac{1 - \pi r^2}{2r}\right)^2}\right) \right).
    \label{eq:psb_lambda}
\end{align}
The result is shown in Fig.~\ref{fig:psb}(a), compared with the numerical value of $\lambda$ using the DPH framework as well as with the result of computing  the term $\Braket{\log(a)}$ in the toy model numerically from the evolution of the perturbation vector norm. All three curves are in excellent agreement for small and intermediate $r$, only for large $r$ does Eq.~\eqref{eq:psb_lambda} slightly deviate from the numerical values because the approximation for $b(r)$  in Eq.~\eqref{eq:b_factor} and thus for $\Braket{\log(a)}$ becomes less accurate.

\subsection{Magnetic periodic Sinai billiard}
We now want to apply the same process to the MPSB, which, however, has a mixed phase space: there exist collision-less orbits like those seen in Fig.~\ref{fig:billiards}(b) that constitute the regular part of phase space (other unstable periodic orbits of zero measure are not relevant here). We are of course only considering the Lyapunov exponent of the chaotic part of the phase space, which means that we initialise particles only in the chaotic phase space region. The mean collision time $\kappa_\text{MPSB}$ between successive collisions with discs is also only defined for the chaotic phase space part (as the regular trajectories do not collide with the discs).

\begin{figure}[t!]
    \centering
    \includegraphics[width=\columnwidth]{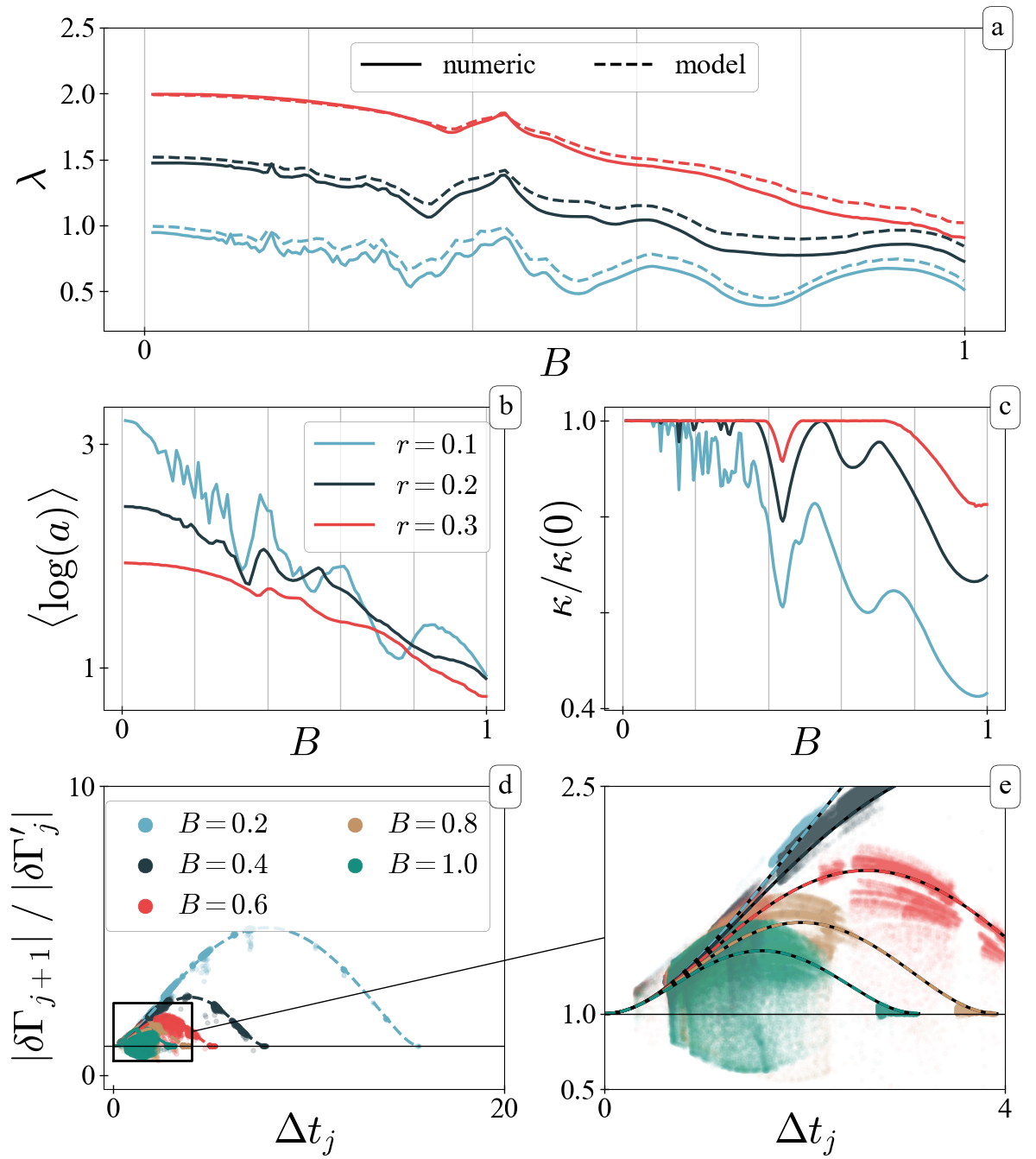}
    \caption{(a) Lyapunov exponent of the magnetic periodic Sinai billiard (MPSB) for different radii versus the magnetic field, compared with the toy model. The solid curves are the numeric result $\lambda$, the dashed curves are the toy model using the numeric average of $\Braket{\log(a)}$. (b) Numeric average of $\Braket{\log(a)}$ versus the magnetic field. (c) Chaotic phase space portion ($g_c = \kappa /\kappa(0)$) of the MPSB ((a-c) share legend). (d, e) Perturbation norm change during the free flight in the MPSB (calculated with the DPH framework). Plotted with dashed lines are Eq.~\eqref{eq:z_MPSB} (data for $r = 0.2$).}
    \label{fig:mpsb}
\end{figure}

The free flight evolution in the MPSB is fundamentally different from the PSB. Not only are the functional forms different but in addition due to magnetic focusing it is possible (and in fact quite frequent) for the perturbation norm to \emph{decrease} during the evolution, as can be seen in Fig.~\ref{fig:mpsb}(e,f). In addition, as visible in Fig.~\ref{fig:pertgrowth}(d), it is also possible for the norm to decrease during the instantaneous change as well. 

This more complex behaviour is of course hidden in the more complicated formulas of our extension to the DPH framework for magnetic fields.  For example, explicitly writing out Eq.~\eqref{eq:evolution_magnetic} gives
\begin{align}
|\dG(t)|  = & \Big[ (\delta p_x^2 + \delta p_y^2) + \nonumber \\
& \left(\delta q_x + \delta p_y\frac{\cos(\omega t) - 1}{\omega} + \delta p_x \frac{\sin(\omega t)}{\omega}\right)^2 + \nonumber \\
& \left. \left(\delta q_y - \delta p_x\frac{\cos(\omega t) - 1}{\omega} + \delta p_y \frac{\sin(\omega t)}{\omega}\right)^2
\right]^{1/2}
\label{eq:B_increase}
\end{align}
(where again we assumed $|\dG(0)|=1$). The consequences of Eq.~\eqref{eq:B_increase} can be seen in Fig.~\ref{fig:mpsb}(d, e). Using a uni-variate scalar function $z(\Delta t)$ to approximate these distributions appears to be a bold move, but in the end it will turn out to give a good approximation. To obtain $z(\Delta t)$ we simplify Eq.~\eqref{eq:B_increase} to 
\begin{equation}
    z_\text{MPSB}(t) = \sqrt{1 + \left(\frac{1 - \cos(\omega t)}{\omega}\right)^2 +  \left(\frac{\sin(\omega t)}{\omega}\right)^2 }.
    \label{eq:z_MPSB}
\end{equation}
which is also plotted in Fig.~\ref{fig:mpsb}(d, e).

In the next step we compute  $\Braket{\log(a)}$ numerically and use its value in the toy model along with $z_\text{MPSB}(\kappa(B))$. We remind that the value of $\kappa$, the mean collision time between discs in MPSB, is not known analytically but it is connected with the chaotic phase space portion through Eq.~\eqref{eq:datseris_proof}. The results of the toy model are presented in Fig.~\ref{fig:mpsb}.

Besides the fact that our toy model approximates $\lambda$ very accurately, Fig.~\ref{fig:mpsb} nicely shows the impact of phase space restriction on $\lambda$. In our toy model the value of $\lambda$ is composed of five contributions, the first being the denominator $\kappa$. The value of $\Braket{\log(a)}$ itself has two contributions, one again stemming from $\kappa$ (as shown in the previous section) and the other from $B$. The function $z_\text{MPSB}$ also has two contributions, one from $B$ and one from $\kappa$. Therefore three out of five contributions to $\lambda$ are inherently linked to the restriction of the chaotic phase space by regular orbits.

\subsection{Mushroom billiard}
\label{sec:mushroom}
\begin{figure}[t!]
    \centering
    \includegraphics[width=\columnwidth]{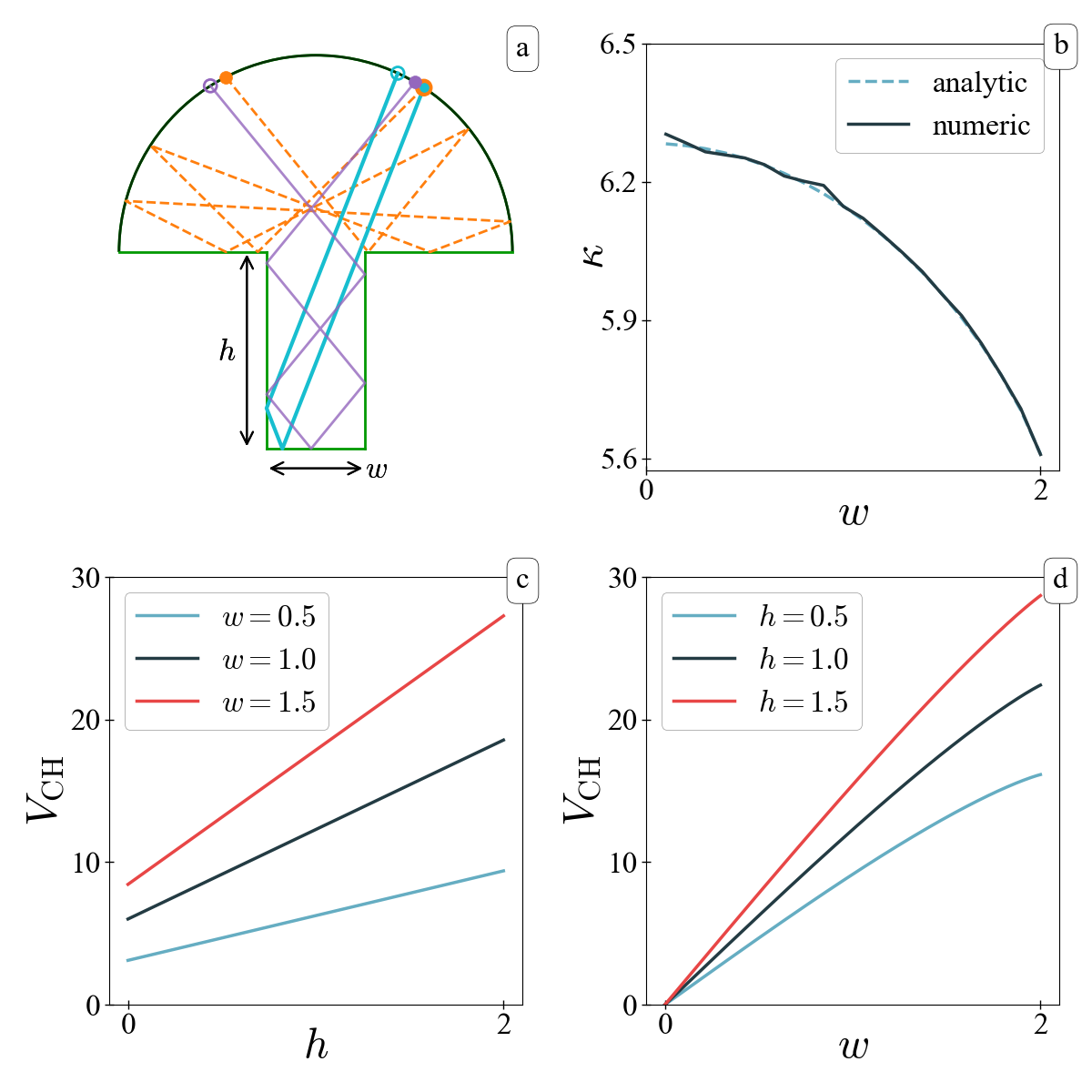}
    \caption{(a) A laminar episode (orange, dashed) and two chaotic episodes (blue, purple) in the mushroom billiard. Starts and ends of each episode are denoted with closed and open circles and the blue episode starts directly after the orange. The cap head is plotted in dark color to differentiate. (b) Mean return time to the stem bottom, which is equivalent with the average elementary growth segment time, eqs.~\eqref{eq:kappa_mb_numeric}, \eqref{eq:kappa_mb_analytic}. (c,d) Volume of chaotic phase space for the mushroom billiard.
    }
    \label{fig:mushroom1}
\end{figure}

Because the volume fractions of the regular and the chaotic phase space regions are not known analytically in the MPSB we have turned to a billiard that also has a mixed phase space but allows to calculate these fractions, and, as we will show, the relevant average time scales analytically: the mushroom billiard (MB). The regular orbits in the MB are orbits forever staying in the cap, evolving exactly like they would as if they were in a circular billiard~\cite{bunimovich_mushrooms_2001}. The tangential circle to these orbits has a radius $\ge w/2$ as shown in Fig.~\ref{fig:billiards}(a). The rest of the orbits, which do not satisfy this criterion, eventually enter the stem and are chaotic. The tangential circle argument was used in~\cite{barnett_quantum_2007} to obtain an analytic expression for the regular phase space volume $V_\text{REG}$ of the MB as a function of the billiard parameters
\begin{align}
  V_\text{REG} &= 2\pi\left( \arccos(\frac{w}{2}) - \frac{w}{2} \left( 1-\frac{w^2}{4} \right) \right) \label{eq:mb_regular} \\
  V_\text{TOT} & =  2\pi (h w + \pi/2), \\
  \label{eq:mb_total}
  V_\text{CH} & = V_\text{TOT} - V_\text{REG}
\end{align}
where $V_\text{TOT}, V_\text{CH}$ is the total and chaotic phase space volume (all lengths are scaled to the cap radius $r$ which is fixed to $r=1$). The parameter dependence of $V_\text{CH}$ is illustraded in Fig.~\ref{fig:mushroom1} c and d. Interestingly, $V_\text{CH}$ does not vanish for small $h$, although it is obvious that there are no chaotic orbits for $h=0$.
This discontinuity is due to the fact that the volume of chaotic phase space in the cap is independent of stem height for nonzero $h$, but drops to zero for $h=0$.

\begin{figure*}[t!]
    \centering
    \includegraphics[width=\textwidth]{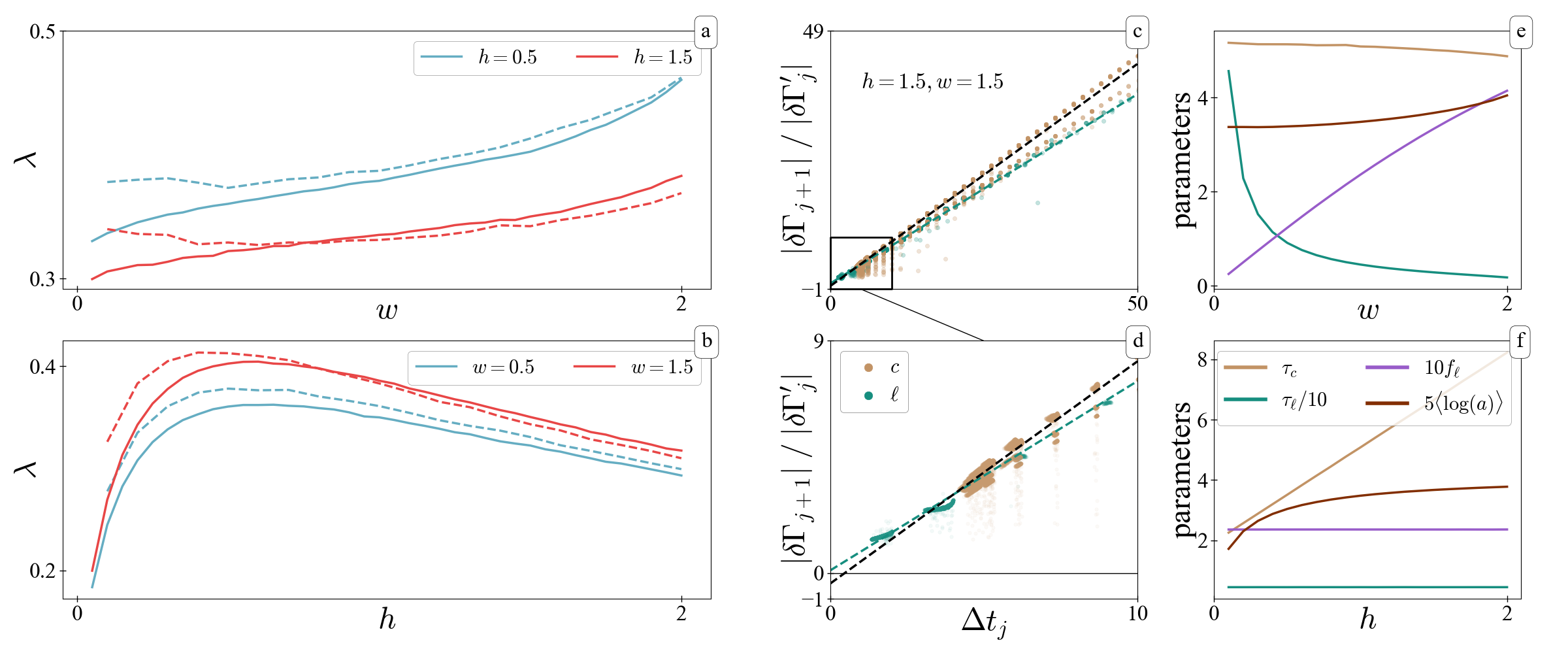}
    \caption{(a, b) Lyapunov exponents in the mushroom billiard (MB) versus the width $w$ or height $h$ of the stem. Solid lines are numeric results using the DPH framework, dashed lines are using the toy model. (c, d) Perturbation norm increase during the chaotic $c$ and laminar $\ell$ episodes. (e, f) Parameters of the toy model versus $w$ or $h$ (for constant $h=1$ and $w=1$ respectively); legend is shared. 
    }
    \label{fig:mushroom2}
\end{figure*}

In Fig.~\ref{fig:mushroom2}(c,d) we present a scatter plot of various possible increases of the perturbation norm during the unit cells. We found that there are clearly distinct contributions to the increase, each seemingly approximated as linear function of $\Delta t$.
By analysing the dynamics in more detail, it turns out that the different contributions of Fig.~\ref{fig:mushroom2}(c,d) come from the trapping of the chaotic orbits in the regular phase space. 
In coordinate space this means that the particles get trapped in the cap and mimic the motion of the regular phase space there until eventually escaping after some time.
This effect is often called \emph{intermittency}, and is known to occur in mushroom billiards~\cite{Altmann2005, Altmann2006}. Intermittent behaviour in the MB can happen in the stem as well, where orbits stay trapped bouncing between the stem walls. 

We therefore have to separate the elementary growth segment into two different ``episodes'': the chaotic episode $c$ and the laminar episode $\ell$, where the particle is trapped in the cap. Notice that the second intermittent behaviour, trapping in the stem, does not lead to a new type of dynamics, but is just prolonged chaotic episodes (similarly to a large free flight in the PSB).
We show the two episodes in Fig.~\ref{fig:mushroom1}(a).
Numeric calculations shown in Fig.~\ref{fig:mushroom2}(e, f) show that each episode has a different average time, $\tau_c, \tau_\ell$ respectively.

During the chaotic episodes the picture is very similar to the PSB. A collision with the cap head gives an instantaneous increase to $|\dG|$, followed by an approximately linear increase until the next collision with the cap head. Here the linear increase approximation is valid because for the chaotic episodes $\Delta t \gtrsim 2h + 2 - w$. After colliding with the cap head the particle may return to the stem immediately which initialises another chaotic episode. Occasionally, after ending a chaotic episode, the particle will get trapped in the cap (see Fig.~\ref{fig:mushroom1}(a) orange), starting a laminar episode. Even though there are successive collisions with the cap head in this episode, the perturbations do not increase exponentially. The successive instantaneous increases are very quickly becoming insignificant (see Fig.~\ref{fig:pertgrowth}(e,f)) due to the fact that cap collisions have an initially focusing effect which only becomes de-focusing if the consecutive free motion is long enough, which is not the case in the laminar episodes. Therefore the overall perturbation growth inside the cap trapping episodes is \emph{linear}.

Let $n_c$ and $n_\ell$ be the counts of chaotic and laminar episodes up to time $T$. Notice that $n_\ell$ is strictly less than $n_c$ since a chaotic episode always follows a laminar episode, but the inverse is only occasionally true. In the limit $T\to\infty$ we define $f_\ell = n_\ell/(n_\ell + n_c)$ to be the frequency of the laminar episodes. We then write the function $z$ as 
\begin{align}
    z_\text{MB}(\Delta t) = (o_c + s_c\Delta t_c) + f_\ell (o_\ell + s_\ell \Delta t_\ell) 
    \label{eq:z_MB}
\end{align}
with $o_i$ the offset and $s_i$ the slope of the linear approximation (we obtain these values with least squares fit to Fig.~\ref{fig:mushroom2}(c, d)). For the chaotic episodes $s, o$ are constant versus $h, w$ while for the laminar episodes $o$ depends strongly on $w$. Also, for the chaotic episodes $o$ has negative value (of around -0.8) which is expected due to the focusing effect. We once again compute $\Braket{\log(z(\Dt))}$ simply by replacing $\Dt$ by its average values $\tau_c, \tau_\ell$ in Eq.~\eqref{eq:z_MB}.

The instantaneous change factor $a$ is the same between the laminar and chaotic episodes so we do not need to separate it. Notice that for the laminar episode we only consider the first jump as the instantaneous increase. Subsequent jumps that decrease rapidly are encoded in the linear growth approximation. After computing $\Braket{\log(a)}$ numerically, we still need a value for $\kappa_\text{MB}$, the elementary growth segment average time, to apply our toy model $\lambda = (\Braket{\log(a)} + \log(z_\text{MB}(\tau_c,\tau_\ell))/\kappa_\text{MB}$. Numerically we can estimate
\begin{equation}
    \kappa_\text{MB} = \tau_c + f_\ell\tau_\ell.
    \label{eq:kappa_mb_numeric}
\end{equation}
However we can estimate $\kappa_\text{MB}$ analytically as well, using Kac's lemma~\cite{kac_notion_1947, mackay_transport_1994,meiss_average_1997}. The key to this is understanding that the mean elementary growth segment time is equivalent with the mean return time to the stem bottom, since all phases in the end of the day have to go there, since all phases are part of the chaotic phase space.

We present the full proof in appendix~\ref{ap:stem}. The final expression is given by
\begin{equation}
\kappa_\text{MB}(h, w) = \frac{V_\text{CH}(h, w)}{2w}
\label{eq:kappa_mb_analytic}
\end{equation}
We compare the analytic formula with the numeric result in Fig.~\ref{fig:mushroom1}(b) and find the expected perfect agreement, since Eq.~\eqref{eq:kappa_mb_analytic} is exact. In appendix~\ref{ap:tauc} we also present an analytic approximation for $\tau_c$. Since we know $\kappa_\text{MB}$ and $\tau_c$ analytically, we also know the product $f_\ell\tau_\ell$ (but we don't have an expression for $f_\ell$ or $\tau_\ell$ individually).

We can now use our toy model to compare with $\lambda$, which we do in Fig.~\ref{fig:mushroom2}(a, b). Again we find good agreement between toy model and the numerical simulation using the DPH framework. The model mildly diverges for very small $w$, probably because the mean laminar time $\tau_\ell$ diverges as seen in Fig.~\ref{fig:mushroom2}(e).

As was the case in the MPSB, the average elementary growth segment time $\kappa$ is inversely proportional to $\lambda$ and directly proportional to the chaotic phase space volume $V_\text{CH}$. This shows that phase space restrictions have an immediate impact in the value of the Lyapunov exponent even for billiards with intermittent dynamics.
Notice that in the MB both $V_\textbf{CH}$ and $\lambda$ increase as $w$ increases. This is simply due to the dependence of $V_\text{CH}$ on $w$, as well as the direct dependence of $\kappa_\text{MB}$ on $1/w$ (this for example was not the case in Eq.~\eqref{eq:datseris_proof} for the MPSB).

\section{Discussion}
To summarise, we have examined the value of the Lyapunov exponent $\lambda$ in chaotic billiards. We were able to create a conceptually simple model that approximates $\lambda$ very well. The model is based on how perturbations evolve in billiards \emph{on average} and helps to understand how each part of the dynamics contributes to the perturbation increase. The simple model is written as eq.~\eqref{eq:toymodel}, which is
\begin{equation}
    \lambda = \frac{1}{\kappa}\left( \Braket{\log\left(a\right)} + \Braket{\log\left(z(\Delta t) \right)} \right) \nonumber
\end{equation}
where $a$ the instantaneous change of $|\dG|$ at a collision with a curved boundary and $z(t)$ the continuous change of $|\dG|$ in between collisions with curved boundaries. $\kappa$ is the average elementary growth segment time equal to the mean collision time between curved boundaries. The approximations that lead to the toy model were the following. First we assumed that the chaotic phase space is ergodic and time averages can be replaced by phase space averages and that for  $\Dt, \log(a), \log(z(\Dt)),$ their averages are finite and greater than 0. We then made the simplifying assumption that the norm of the perturbation vector increases continuously in between successive chaos-inducing collisions (i.e. in each elementary growth segment) as a \emph{univariate} function of the time interval $z(\Dt)$. 

\begin{figure}
    \centering
    \includegraphics[width=\columnwidth]{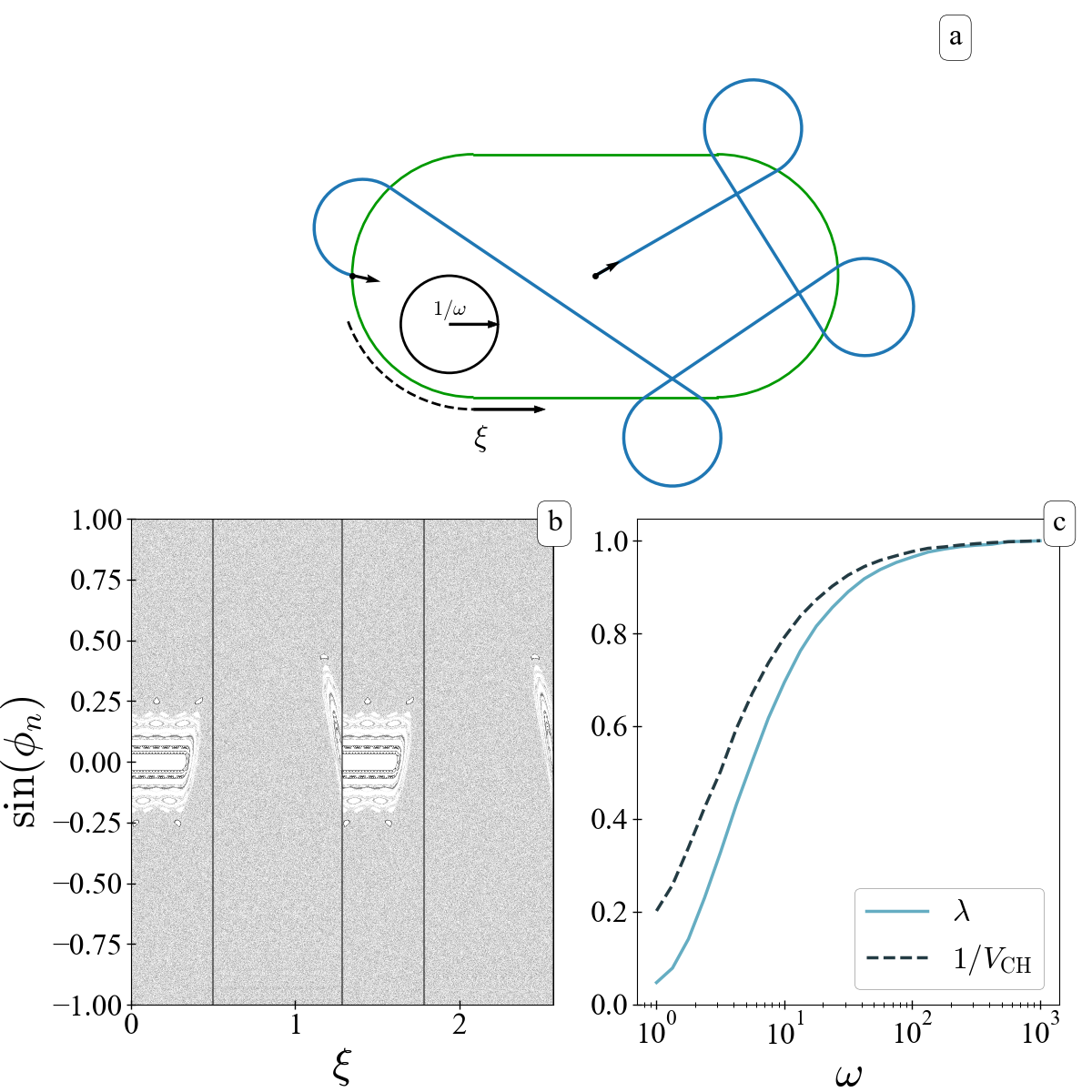}
    \caption{The inverse stadium billiard (ISB). (a) An example orbit in the ISB (stadium width and length are 0.5). Outside the stadium the particle undergoes circular motion with radius $1/\omega$. (b) Boundary map (see Appendix \ref{ap:stem}) of the ISB, computed for $\omega = 10$. In the middle one can see stability islands, which seem to have a fractal boundary. (c) Lyapunov exponent $\lambda$ and chaotic volume $V_text{CH}$ versus $\omega$, both normalized to their maximums for comparison. $\lambda$ is obtained with a modified version of the DPH framework using tangent space evolution matrices derived by V\"or\"os et al. in~\cite{voros_tunable_2003}. $V_\text{CH}$ is the volume of the billiard flow of the chaotic orbits and is calculated by weighting the area of the ergodic region (obtained numerically) on the boundary map with its mean time to next collision. We only consider orbits that do intersect the billiard boundary of the ISB.}
    \label{fig:isb}
\end{figure}

We used Eq.~\eqref{eq:toymodel} to find an analytic expression for $\lambda$ in the periodic Sinai billiard (PSB). 
We have also shown that  Eq.~\eqref{eq:toymodel} can be used to analyse the Lyapunov exponent in the magnetic periodic Sinai billiard (MPSB), and by approximating the numerical curves identified the main contributions.
We could follow the same approach for the mushroom billiard (MB), even though the process is complicated in this case by intermittent dynamics. 
In both billiards with mixed phase space we connected the chaotic phase space volume with $\lambda$ through $\kappa$ and showed that $\lambda$ has a leading contribution given by the inverse of the chaotic phase space volume (for the MPSB we used the chaotic \emph{portion} instead of volume, because the total phase space volume does not depend on $B$).

To strengthen our point that a prominent contribution to the parameter dependence of the Lyapunov exponent in billiard is given by the inverse chaotic phase space volume, we present one final billiard, called the inverse stadium billiard (ISB) shown in Fig.~\ref{fig:isb}(a), originally studied by V\"or\"os et al. in Ref.~\cite{voros_tunable_2003}. In this billiard a particle is propagating inside the stadium on straight lines, but after passing the boundary of the stadium it is subjected to a constant magnetic field, which brings the particle back inside the stadium, as depicted in Fig.~\ref{fig:isb}(a). In the limit of infinite magnetic field the ISB recovers the fully chaotic Bunimovich stadium, for finite magnetic fields it has a mixed phase space. Here, we don't want to analyze  the ISB in any detail but only point out that also in this billiard the parameter dependence of its Lyapunov exponent is closely following the inverse of the chaotic phase space volume as shown in Fig.~\ref{fig:isb}(b,c).

We want stress how different the mechanisms are that lead to chaos in the three different billiards. In the MPSB it is dispersing as well as the magnetic field. In MB it is defocusing and in the ISB it is even more involved. Yet in all three cases we find what is suggested by our toy model: the Lyapunov exponent has a leading contribution that is inverse to the chaotic phase space volume. 

Because generally chaos in billiards arises via dispersing and defocusing collisions with curved boundary segments~\cite{Bunimovich2018}, the Lyapunov exponent is necessarily inversely linked with the mean return time to these boundary segments.
Furthermore, Kac's lemma dictates that the mean return time of the chaotic trajectories to these boundaries is directly proportional to the chaotic phase space volume.
For this reason we hypothesise that for most chaotic billiards with mixed phase space the Lyapunov exponent has a leading contribution inverse to the chaotic phase space volume.

This should straight-forwardly carry over to higher dimensions as well, since Kac's lemma, the DPH framework, as well as our toy model, do not depend in any way on the dimensionality of the billiard. 
So far we didn't find significant differences between billiards with sharply divided phase space (like the MB and the MPSB) and a fractal phase space structure (like the ISB). To conclude whether there are fundamental differences between sharply-divided and fractal phase spaces one will have to do more research. What we want to point out is that for fractal phase spaces it is much harder to estimate the volume of the chaotic set.

\appendix

\section{Evolution of perturbation vector in a magnetic field}
\label{ap:magnetic}
In their paper, Dellago, Posch and Hoover give two main equations to compute the evolution of perturbations in tangent space along a piece-wise smooth flow defined by the autonomous ODE system
\begin{equation}
    \dot{\vec{\Gamma}} = \vec{F}(\vec{\Gamma}).
\end{equation}
During smooth propagation, the perturbation vector $\dG$ along the trajectory $\vec{\Gamma}$ evolves according to 
\begin{equation}
    \label{eq:dph_evo_smooth}
    \dot{\dG} = \left.\pdv{\vec{F}}{\vec{\Gamma}}\right|_\vec{\Gamma} \cdot \dG.
\end{equation}
If smooth propagation is interrupted at discrete times $t_j(\vec{\Gamma})$ by a discontinuous jump, represented here by a differentiable map $\vec{M}_j(\vec{\Gamma})$, the perturbation vector after the jump is given by
\begin{equation}
    \label{eq:dph_evo_jump}
    \begin{aligned}
        \dG' = &\eval{\pdv{\vec{M}_j}{\vec{\Gamma}}}_{\vec{\Gamma}} \cdot\,\dG_i\,+ \\ &\left(\eval{\pdv{\vec{M}_j}{\vec{\Gamma}}}_{\vec{\Gamma}}\cdot\vec{F}(\vec{\Gamma}_i)-\vec{F}\qty(\big.\vec{M}(\vec{\Gamma}))\right)\delta\tau_c
    \end{aligned}         
\end{equation}
where $\delta\tau_c = t_j(\vec{\Gamma} + \dG) - t_j(\vec{\Gamma})$.
For the case of elastic reflection with an obstacle, $\vec{M}_j$ can  be written as
\begin{equation}
  \label{eq:dph_map_elastic}
  \vec{\Gamma}' = \vec{M}_j(\vec{\Gamma}) = \left(\begin{array}{c|c}
    \mathbb{I}_{2\times2} & 0_{2\times2}\\\hline
    0_{2\times2} & \mathbb{I}_{2\times2} - 2(\vec{n}\otimes\vec{n}) 
  \end{array}\right)\cdot\vec{\Gamma}.
\end{equation}
Here $\vec{n}$ is the normal vector of the obstacle at the collision point, $\mathbb{I}_{2\times2}$ and $0_{2\times2}$ are the $2\times2$ identity and zero matrices respectively, and $\qty(\vec{a}\otimes \vec{b})_{jk} = a_jb_k $ is a second-order tensor.

\subsection{Propagation}
In magnetic billiards, particles propagate in circular arcs. 
To get the simplest possible set of equations of motion describing this mode of propagation, it is useful to introduce a phase angle $\theta = \arctan(p_y/p_x)$.
For uniform circular motion, this phase angle grows linearly in time as the system rotates with constant angular velocity $\omega$. 
This can be exploited to determine $\dot{\vec{p}}$ using the chain rule
\begin{align}
  \label{eq:ftls_dpi}
  \dot{\vec{p}} = \dv{\vec{p}}{t} = \dv{\theta}{t}\cdot\dv{\vec{p}}{\theta} =
  \omega\cdot\dv{\vec{p}}{\theta}.
\end{align}
Using $\lVert\vec{p}\rVert = 1$, one can easily calculate the explicit relation between $\vec{p}$ and $\theta$
\begin{align}
  \label{eq:ftls_dptheta}
  \begin{aligned}
    p_x &= \cos(\theta)\\
    p_y &= \sin(\theta)
  \end{aligned}
          \hspace{1.5em}\Rightarrow\hspace{1.5em}
  \begin{aligned}
    \dv{p_x}{\theta} &= -p_y\\
    \dv{p_y}{\theta} &= p_x
  \end{aligned} 
\end{align}
and combine the results of eqs. \eqref{eq:ftls_dpi} and \eqref{eq:ftls_dptheta} to
receive the equations of motion

\begin{align}
  \label{eq:dph_mag_eom}
  \vec{F}(\vec{\Gamma}) = \pmqty{\vec{p}\\\omega\cdot\vec{R}\,\vec{p}}
  \quad\text{where}\;\vec{R} = \spmqty{0 & -1\\  1 &  0}.
\end{align}
Using Eq.~\eqref{eq:dph_evo_smooth}, we can now state the equations of
evolution for a perturbation vector $\dG$ using the Jacobian $\vec{J}$ of $\vec{F}$.
\begin{align}
  \label{eq:ftls_jacobian}
  \dot{\dG} = \vec{J} \cdot\dG = \left(
  \begin{array}{c|c}
    0_{2\times2} & \mathbb{I}_{2\times2}\\ \hline
    0_{2\times2} & -\omega \cdot \vec{R} 
  \end{array}
              \right)\dG.
\end{align}
Using an exponential ansatz, we can compute the general solution of this system and receive a final result of
\begin{align}
  & \dG(t) = \mathbb{B} \cdot \dG(t_0), 
  \label{eq:dph_mag_smooth}\\
  & \mathbb{B} = \left(
  \begin{array}{c|cc}
    \mathbb{I}_{2\times2} & 
             \begin{array}{c}
               \rho \sin(\omega t)\\
               -\rho \left(\cos(\omega t) - 1\right)
             \end{array} &
                           \begin{array}{c}
                             \rho \left(\cos(\omega t) - 1\right)\\
                             \rho \sin(\omega t) 
                           \end{array} \\ \hline
    0_{2\times2}&
             \begin{array}{c}
               \cos(\omega t)\\
               \sin(\omega t)
             \end{array} &
                           \begin{array}{c}
                             -\sin(\omega t)\\
                             \cos(\omega t)
                           \end{array}
  \end{array}
  \right) \nonumber
\end{align}
where $\rho = 1/\omega$ is the cyclotron radius.
\subsection{Collisions}
The derivation of the collision map for $\dG$ is largely analogous to the process used by  DPH to derive their result for non-magnetic billiards.
The equations of motion are assumed as stated in Eq. \eqref{eq:dph_mag_eom}.
For equation \eqref{eq:dph_evo_jump}, we require the Jacobian matrix of $\vec{M}_j$, which is
\begin{equation}
  \label{eq:dph_mag_jacobian}
  \frac{\partial\vec{M}_j}{\partial\vec{\Gamma}} =
  \begin{pmatrix}
    \mathbb{I}_{2\times2} & 0\\
    \vec{A} & \vec{B}
  \end{pmatrix}
\end{equation}
\begin{align*}
  \text{where}\quad
  \vec{A} &= 2\left((\vec{n}\otimes\vec{p}) + \left<\vec{p},\vec{n}\right>·\mathbb{I}_{2\times2}\right){\frac{\partial\vec{n}}{\partial\vec{q}}}\\
  \vec{B} &= \mathbb{I}_{2\times2} - 2\,(\vec{n}\otimes\vec{n}).
\end{align*}
By inserting eqs. \eqref{eq:dph_mag_eom}, \eqref{eq:dph_map_elastic} and \eqref{eq:dph_mag_jacobian} into \eqref{eq:dph_evo_jump}, we find
\begin{equation}
    \begin{aligned}
        \label{eq:dph_mag_monster}
      \dG' = 
      &\begin{pmatrix}
        \mathbb{I}_{2\times2} & 0\\
        \vec{A} & \vec{B}
      \end{pmatrix}  \dG + \left[
        \begin{pmatrix}
          \mathbb{I}_{2\times2} & 0\\
          \vec{A} & \vec{B}
        \end{pmatrix}\right.\cdot\\
        &\left.\begin{pmatrix}
          \vec{p}\\
          \omega\cdot\vec{R}\,\vec{p}
        \end{pmatrix} -
        \begin{pmatrix}
          \vec{p} - 2\,(\vec{n}\otimes\vec{n})\,\vec{p}\\
          \omega\cdot\vec{R}\,\vec{B}\,\vec{p}
        \end{pmatrix}
      \right] \delta\tau_c.
  \end{aligned}
\end{equation}
It is now helpful to continue calculations for $\delta\vec{q}'$ and $\delta\vec{p}'$ separately.
For the position component $\delta\vec{q}'$, equation \eqref{eq:dph_mag_monster} can be written as
\begin{align}
  \delta\vec{q}' &= \delta\vec{q} + \left[ \vec{p} - \vec{p} +
  2\,(\vec{n}\otimes\vec{n})\,\vec{p}\right]\delta\tau_c\nonumber\\
               \label{eq:dph_mag_position}
             &= \delta\vec{q} + 2\,\delta\tau_c\,(\vec{n}\otimes\vec{n})\,\vec{p}.
\end{align}
For the momentum component $\delta\vec{p}'$, we get
\begin{align}
  &\delta\vec{p}' = \vec{A}\,\delta\vec{q} + \vec{B}\,\delta\vec{p} + \delta\tau_c\left[\vec{p}\vec{A} + \omega\mathbb{S}\right]
  \label{eq:dph_mag_momentum}\\
  &\text{where }\mathbb{S}:=\vec{B}\vec{R} - \vec{R}\vec{B}  = 2
  \begin{pmatrix}
    - 2n_1n_2 & n_1^2 - n_2^2\\
    n_1^2-n_2^2 & 2n_1n_2
  \end{pmatrix}\nonumber.
\end{align}
This can be  simplified by using the fact that $(\vec{b} \otimes \vec{a})
\vec{c} = \left<\vec{c}, \vec{a}\right>\vec{b}$ and introducing the quantity
$\delta\vec{q}_c = \delta\vec{q} + \delta\tau_c\,\vec{p}$, which represents the real space
difference vector between the collision points of satellite and reference
trajectories.
\begin{align}
  \label{eq:dph_mag_momentum2}
  \delta\vec{p}' = &\delta\vec{p} - 2 \left<\delta\vec{p}, \vec{n}\right>\vec{n} - 2
  \pdv{\vec{n}}{\vec{q}}\left(\left<\vec{p}, \delta\vec{q}_c\right>\vec{n} +\right.\\
  &\left.\left<\vec{p}, \vec{n}\right>\delta\vec{q}_c\right) + \delta\tau_c\,\omega\,\mathbb{S}\vec{p} \nonumber.
\end{align}
Using geometric considerations outlined by DPH
we can now rewrite the penultimate term to get
\begin{equation}
  \label{eq:dph_mag_momentum3}
  \delta\vec{p}' = \delta\vec{p} - 2 \left<\delta\vec{p}, \vec{n}\right>\vec{n} - 2\gamma_R
  \frac{\left<\delta\vec{q}, \vec{e}\right>}{\cos \phi}\vec{e}' + \delta\tau_c\,\omega\,\mathbb{S}\vec{p}
\end{equation}
where $\phi$ is the angle of incidence, $\gamma_R$ is the local curvature of the
obstacle and $\vec{e}$ and $\vec{e}'$ are unit vectors orthogonal to
$\vec{p}$ and $\vec{p}'$ respectively.
\subsection{Collision delay time}
The quantity $\delta\tau_c$ in eqs. \eqref{eq:dph_evo_jump} and \eqref{eq:dph_mag_monster} can be interpreted as the time delay in between the collisions of the reference trajectory $\vec{\Gamma}$ and its satellite $\vec{\Gamma} + \dG$.

It can be computed by determining the signed distance from the satellite to its collision point measured along its trajectory at the time $t_j(\vecG)$ of the collision of the reference particle.
As we are considering the linearized dynamics of the perturbation and Eq. \eqref{eq:dph_evo_jump} is valid only to the first order of $\dG$, we will ignore all higher orders of $\dG$ in the subsequent calculations.

\begin{figure}[t!]
  \centering
  \includegraphics[width=0.9\linewidth]{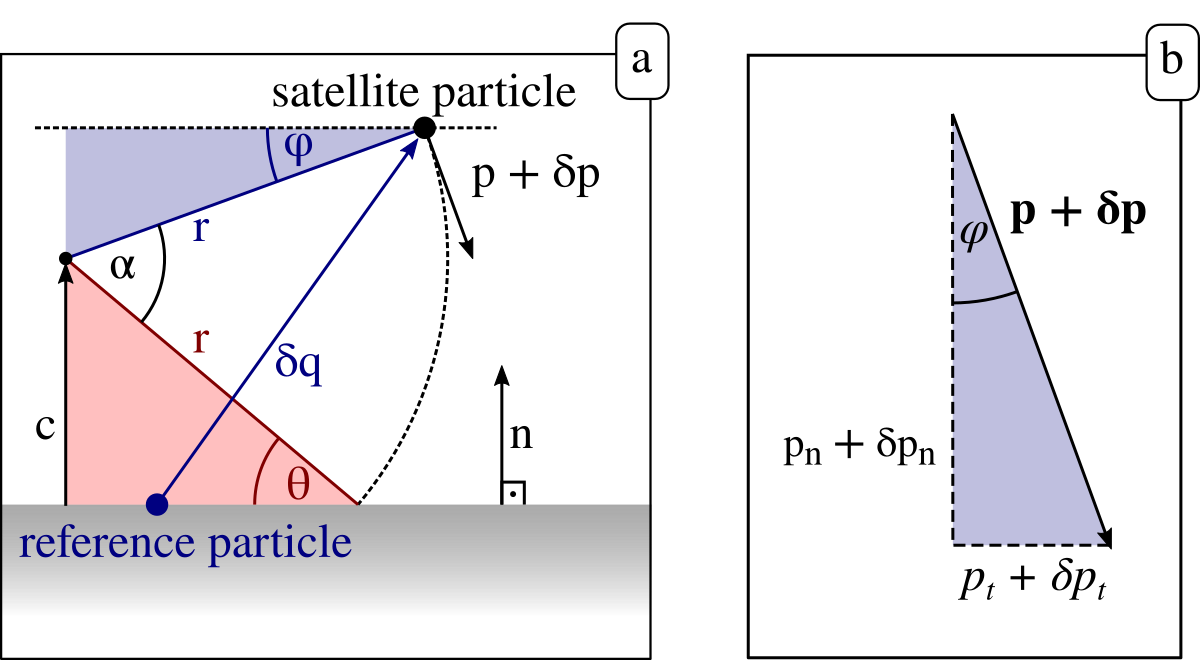}
  \caption{(a) Geometric derivation of $\delta\tau_c$. As all spatial perturbations are
    small, it is sufficient to approximate the obstacle as a straight line. In this example, the satellite particle collides after the reference particle. (b) Decomposition of momentum into normal and tangential components}
  \label{fig:tauc}
\end{figure}
Furthermore, we will denote vector components in normal direction of the obstacle by a subscript $n$, i.e. $a_n = \left<\vec{a}, \vec{n}\right>$. 
Similarly, the tangent component of a vector will be denoted by a subscript $t$.

Geometrically, one can immediately derive the following two relations from the two triangles highlighted in Fig. \ref{fig:tauc}a
\begin{align}
  \rho \sin \phi &= \left<\delta\vec{q},\vec{n}\right> - \left<\vec{c},\vec{n}\right>
  &&\text{(blue triangle)}\\
  \rho \sin \theta &= \left<\vec{c},\vec{n}\right>&&\text{(red triangle)}
\end{align}
where $\vec{c}$ is a vector between the obstacle and the cyclotron centre, as shown in fig. \ref{fig:tauc}a. 
Eliminating the factor $\left<\vec{c},\vec{n}\right>$ yields
\begin{equation}
  \sin \theta =  \frac{\left<\delta\vec{q},\vec{n}\right>}{\rho} - \sin \phi.
  \label{eq:ftls_sintheta}
\end{equation}
By construction, $\theta$ cannot exceed $\frac{\pi}{2}$ in absolute value.
Therefore, we can safely apply the arcsine function to eq.~\eqref{eq:ftls_sintheta}, receiving an expression for $\theta$.
\begin{equation}
  \label{eq:ftls_theta}
  \theta = \arcsin\left( \frac{\left<\delta\vec{q},\vec{n}\right>}{r} - \sin \phi\right).
\end{equation}
We can now use this to compute the angle $\alpha$ corresponding to the arc the particle has to travel during $\delta\tau_c$.

Using Fig.~\ref{fig:tauc}a, we can determine an expression for $\alpha$
\begin{align}
  \alpha &= \phi + \theta\nonumber\\
    &\overset{\mathclap{\text{Eq. \eqref{eq:ftls_theta}}}}{=}\hspace*{0.4cm} \phi +
      \arcsin\left( \frac{\left<\delta\vec{q},\vec{n}\right>}{r} - \sin \phi\right)
        \label{eq:ftls_alpha1}.
\end{align}
This can be further simplified by expressing $\phi$ in terms of the satellite particle's momentum (compare fig. \ref{fig:tauc}b).
\begin{align}
  \label{eq:ftls_finaler}
  \alpha &= \underbrace{\arcsin\left(\frac{p_t + \delta p_t}{\norm{\vec{p} + \delta\vec{p}}}\right)}_{K} +\\ 
  &\underbrace{\arcsin  \left(\frac{\left<\delta\vec{q},\vec{n}\right>}{r} - \frac{p_t + \delta p_t}{\norm{\vec{p} + \delta\vec{p}}}\right)}_{L}\nonumber.
\end{align}
We can now linearize Eq. \eqref{eq:ftls_finaler} to receive our final result for $\alpha$. 
As the individual terms are somewhat complicated, but very similar, we will treat them separately.
The first-order Taylor expansion of leftmost term is
\begin{equation}
  \label{eq:ftls_trick1}
  K \approx \arcsin\left(\frac{p_t}{\norm{\vec{p}}}\right) + \zeta(\delta\vec{p_t})
\end{equation}
where $\zeta$ is of the form $a\cdot{}\delta p_t + b\cdot{}\delta p_n$ with $a,b$ given by the respective partial derivatives. 
Expanding the rightmost term in \eqref{eq:ftls_finaler} yields the similar result of
\begin{equation}
  \label{eq:ftls_trick2}
  L\approx\arcsin\left(-\frac{p_t}{\norm{\vec{p}}}\right) - \zeta(\delta\vec{p_t}) +
  \frac{1}{\sqrt{1 - p_t^2}}\cdot\frac{\left<\delta\vec{q_i},\vec{n}\right>}{r}.
\end{equation}
This simplifies the final result for $\alpha$ significantly. 
Using the anti-symmetry of the arcsine function, we can see that most of \eqref{eq:ftls_trick1} and \eqref{eq:ftls_trick2} cancel out, leaving only
\begin{equation}
  \label{eq:ftls_tada}
  \alpha = \frac{1}{\sqrt{1 - p_t^2}}\cdot\frac{\left<\delta\vec{q},\vec{n}\right>}{r}.
\end{equation}
Finally, we have to multiply equation \eqref{eq:ftls_tada} with the cyclotron radius $r$, then divide the resulting arclength by $\norm{p}$ to get $\delta\tau_c$.

As $\norm{\vec{p}} = 1$ per convention, we can substitute $\sqrt{1 - p_t^2} = \abs{p_n}$. However, we know that $p_t < 0$ as the reference particle must have been moving towards the obstacle to collide with it.
Therefore, we can further simplify our result to obtain
\begin{equation}
  \label{eq:ftls_theveryend}
  \delta\tau_c = - \frac{\left<\delta\vec{q},\vec{n}\right>}{\left<\vec{p},\vec{n}\right>}.
\end{equation}
This is the same result as for linear propagation in~\cite[Eq.~18]{dellago_lyapunov_1996}, since higher orders of $\dG$ were neglected.

Combining eqs. \eqref{eq:dph_mag_position}, \eqref{eq:dph_mag_momentum3} and \eqref{eq:ftls_theveryend}, we receive the final result
\begin{align}
    \dG'  = &\pmqty{
    \delta\vec{q} - 2\left(\delta\vec{q}\cdot\vec{n}\right)\vec{n}\\
    \delta\vec{p} - 2\left(\delta\vec{p}\cdot\vec{n}\right)\vec{n} - 2\gamma_r\;
    \tfrac{\left<\delta\vec{q},\vec{e}\right>}{\cos \phi}\;\vec{e}'\\
  } - \nonumber \\
  &\omega\,\frac{\left(\delta\vec{q}\cdot\vec{n}\right)}{\left(\vec{p}\cdot\vec{n}\right)}
  \pmqty{0\\ \mathbb{S}\cdot \vec{p}}
  \label{eq:dph_mag_final}.
\end{align}

\section{Mean return time to stem}
\label{ap:stem}

In this section, we will derive an analytic expression for the mean return time to the stem bottom in a mushroom billiard. 
Using Kac's lemma~\cite{kac_notion_1947, mackay_transport_1994,meiss_average_1997}, which states that for volume-preserving maps, the mean number of iterations $n_S$ required to return to a compact subset $S$ of phase space is given by
\begin{equation}
    \label{eq:kac}
    n_S = \frac{\mu(A)}{\mu(S)}
\end{equation}
where $\mu(\cdot)$ is the volume of a set and $A$ is the subset of phase space accessible to orbits originating in $S$.

To transform the billiard flow into a map, we discretize time in small steps $\Delta t$, implicitly considering the limit of $\Delta t \rightarrow 0$. We now choose the set $S$ of momenta and positions defined by 
\begin{equation}
    S = \{\vec{q}\in Y, p_y > 0\}
\end{equation}
where $Y$ is a box of width $w$ and height $\epsilon$ at the bottom of the stem. One should be careful about the choice of $S$. The simplistic approach of choosing the cap semicircle as the returning set (since the elementary growth segments are delimited by collisions with curved boundaries) will not yield the correct result. That is because the mean return time to the cap semicircle inherently includes contributions from both the periodic orbits of the MB as well as the laminar episodes of the elementary growth segments, which we have already shown to correspond to ``free flight''-like motion.

The phase space volume of $S$ is $\mu(S) = \pi w \epsilon$ in the limit of $\Delta t\rightarrow 0$.
As the chaotic phase space component of mushroom billiards is ergodic~\cite{bunimovich_mushrooms_2001},
we know that the measure of the subset $A$ of phase space accessible from $S$ is given by 
the volume of chaotic phase space $V_\text{CH} = V_\text{TOT} - V_\text{REG}$.

Applying Kac's lemma to get the mean iterations to return to the stem and multiplying by $\Delta t$, we 
get a result for the mean return time $\kappa_S(h, w, r)$ to the set $S$.
\begin{equation}
    \kappa_S(h,w) = \frac{V_\text{CH}(h,w)}{\pi \epsilon w} \Delta t.
\end{equation}
To eliminate $\Delta t$, we divide by $\kappa_S$ for $w = 2$ to get
\begin{equation}
    \label{eq:kappa_prop}
    \kappa_S(h,w) = \frac{V_\text{CH}(h,w)}{V_\text{TOT}(h, 2)}\frac{2}{w} \kappa_S(h, 2).
\end{equation}
As this equation no longer depends on $\epsilon$, we can now take the limit $\epsilon \rightarrow 0$.
The reason to use $w=2$ here is because the MB becomes the stadium billiard for $w=2$ (and thus is fully chaotic with no regular components with measure $> 0$).


We can find $\kappa_S(h, 2)$ because of the ergodicity of the MB for $w=2$. Specifically, it holds that $\kappa_S(h, 2) = n_S(h)\times \tau$ with $n_S$ the mean amount of iterations to return to the stem and $\tau$ the mean collision time in the MB given by Eq.~\eqref{eq:universal}.
We consider the boundary map of the billiard (Birkhoff coordinates), $(\xi, \sin{\phi_n})$, with $\xi$ the coordinate along the boundary (i.e. the arc-length) and $\phi_n$ the angle of incidence with respect to the normal vector at $\xi$.
This coordinate system is a discrete mapping and Kac's lemma applies directly.
Therefore the mean iterations to return to the stem bottom are
\begin{equation}
    \label{eq:ns_mb_ergodic}
    n_S = \frac{2\abs{\partial Q}}{2\cdot 2}
\end{equation}
where $\abs{\partial Q}$ is the perimeter the boundary (the explicit factor of 2 represents the contribution of $\sin\phi_n$).
Using Eq.~\eqref{eq:universal}, we find
\begin{equation}
    \label{eq:tau_mb_ergodic}
    \tau = \frac{\pi\cdot (2h + \pi/2)}{\abs{\partial Q}}.
\end{equation}
Combining eqs. \eqref{eq:ns_mb_ergodic} and \eqref{eq:tau_mb_ergodic}, we receive an expression for the mean stem return time in the fully ergodic case 
\begin{equation}
    \kappa_S(h, 2) = \frac{\pi}{2} (2h + \pi/2).
\end{equation}
This expression can be simplified by substituting the total phase space volume as defined in eq. \eqref{eq:mb_total}, yielding
\begin{equation}
    \kappa_S(h, 2) = \frac{V_\text{TOT}(h, 2)}{4}.
    \end{equation}
Inserting this result into eq. \eqref{eq:kappa_prop}, we receive
\begin{equation}
    \kappa_\text{MB}(h, w) = \frac{V_\text{CH}(h, w)}{2w}.
    \label{eq:kappa_mb_appendix}
\end{equation}  

\section{Mean duration of chaotic episodes}
\label{ap:tauc}
A chaotic episode in the MB as defined above consists of the particle travelling from the cap head directly into to the stem and back up to the cap head, without any other collisions inside the cap. 
To determine the mean duration $\tau_c$ of these episodes, we can geometrically determine the average lengths of the trajectories, exploiting that all trajectories are uniquely defined by the angle of incidence $\alpha$ and the distance $\delta$ from the cap center at which they enter the stem. 

To simplify the calculation, it is useful to split the trajectory into the mean length of cap transit $\Braket{c}$ and the mean length of stem transit $\Braket{s}$ where $\tau_c = 2\Braket{c} + 2\Braket{s}$.

The stem transit length can be easily computed using simple trigonometry and depends only on the angle $\alpha$
\begin{equation}
    s = \frac{h}{\cos(\alpha)}.   
\end{equation}
As the directions of particle momenta are equidistributed, we know that $\alpha$ must be cosine-distributed. We can now integrate over $\alpha$ to obtain the mean $s$, finding
\begin{equation}
\Braket{s} = \int \frac{h}{\cos{\alpha}} \frac{1}{2}\cos(\alpha) d\alpha = \pi h .
\end{equation}

Determining the cap transit length is more difficult as it depends both on $\alpha$ and $\delta$. Using the law of sines, we can derive 
\begin{equation}
    c = \frac{r}{\cos(\alpha)} \cos\left(\asin\left(\frac{\delta}{r}\cos{\alpha}\right) 
        - \alpha\right).
\end{equation}
To get an average result, this expression has to be integrated both over $\delta$ and $\alpha$, again using the fact that $\alpha$ is cosine-distributed. 
Unfortunately, we were unable to solve the integrals analytically. 
Therefore, we decided to approximate $\int c\,d\delta$ by a polynomial before performing the second integration, yielding
\begin{equation}
    \Braket{c} \approx 1 - \frac{w^2}{36} - \frac{w^4}{1200} - \cdots\; .
\end{equation}

\section{Proof of Eq.~\eqref{eq:datseris_proof}}
\label{ap:sinaiproof}

We are interested in the flow of the PSB but to apply Kac's lemma we need a discrete system. Thus, we obtain a map of the flow $\Phi^t$ by discretizing in time (similarly with appendix~\ref{ap:stem}), $f := \Phi^{\Delta t_\varepsilon}$, with $\Delta t_\varepsilon \sim \varepsilon$. To prove eq.~\eqref{eq:datseris_proof} we will apply Kac's lemma to a set $\mathcal{S}$.

Let $\mathcal{T}_\mathcal{S} (B) = \Delta t_\varepsilon\times  n_\mathcal{S} (B)$ (where $B$ is the magnetic field) denote physical recurrence time instead of map iterations.
Let $\mathcal{W}$ be a circle of radius $r + \varepsilon$ concentric to the disk of the PSB and define the phase space subset $\mathcal{S}_\varepsilon$ such that
\begin{equation}
\mathcal{S}_\varepsilon = \{\vec{x}, \vec{v}: \vec{x} \in \mathcal{W} \; \text{and} \; \vec{v} \cdot \vec{\eta}(\vec{x}) < 0 \}
\end{equation}
where $\vec{\eta}(\vec{x})$ is the vector normal to $\mathcal{W}$. The mean collision time $\kappa$ of the PBS is exactly $\mathcal{T}_\mathcal{S}$ in the limit $\varepsilon \to 0$.

To find $\mu(\mathcal{S};B)$, the measure of set $\mathcal{S}$, we first realise that it does not depend on the magnetic field, $\mu(\mathcal{S};B) = \mu(\mathcal{S};0) = \mu(\mathcal{S})$.
This is due to the infinitesimal width of $\mathcal{S}$, over which motion can always be approximated by a straight line for all finite magnetic fields values (i.e. equalling the magnetic field free case).
Then, using \eqref{eq:kac} at $B=0$, we have
\begin{equation}
\mathcal{T}_\mathcal{S}(B=0) =\left( \frac{\Delta t_\varepsilon}{\mu(\mathcal{S_\varepsilon})} \right)  \mu(A;B=0) = \left( \frac{\Delta t_\varepsilon}{\mu(\mathcal{S_\varepsilon})} \right),
\label{convergence}
\end{equation}
because the PSB without magnetic field is fully ergodic and thus $\mu(A;B=0)=\mu(M)=1$ (here $M$ denotes the entire phase space whose measure we set for simplicity to 1).
By substitution we get
$
\mathcal{T}_\mathcal{S}(B) = \mu(A;B)\times \mathcal{T}_\mathcal{S}(0)
$.
For small enough magnetic fields all chaotic orbits (and up to measure 0 only those) collide with the disks.
In the limit $\varepsilon\to 0$ both $\Delta t_\varepsilon$ and $\mu(\mathcal{S}_\varepsilon) $ go to 0 linearly with $\varepsilon$, therefore their ratio converges, i.e. $\mathcal{T}_\mathcal{S} \to \kappa$.
Since by definition $\mu(A)=g_c$, the portion of chaotic orbits in the PSB (because we set the entire volume to have measure 1), we find that the mean collision time is given by the fraction of chaotic orbits as a function of the magnetic field $B$, times the mean collision time at $B=0$
\begin{equation}
\kappa(B) = g_c(B)\times  \kappa(0).
\end{equation}

\bibliography{Billiards}
\end{document}